\documentclass[aps, prl, twocolumn, showpacs, 10pt]{revtex4-1}
\usepackage{graphicx, amsmath, amssymb, bm}

\begin{document}

\title{Consensus time and conformity in the adaptive voter model}

\author{Tim Rogers}
\affiliation{Centre for Networks and Collective Behaviour, Department of Mathematical Sciences, University of Bath, Claverton Down, Bath, BA2 7AY, UK}
\author{Thilo Gross}
\affiliation{Department of Engineering Mathematics, University of Bristol, Woodland Road, Bristol, BS8 1UB, UK}

\begin{abstract}
The adaptive voter model is a paradigmatic model in the study of opinion formation. Here we propose an extension for this model, in which conflicts are resolved by obtaining another opinion, and analytically study the time required for consensus to emerge. Our results shed light on the rich phenomenology of both the original and extended adaptive voter models, including a dynamical phase transition in the scaling behavior of the mean time to consensus.
\end{abstract}


\maketitle

In nature, collective intelligence is observed in a wide variety of species. Quite generally groups of animals are able to aggregate information and make decisions jointly \cite{Couzin2011}. The most impressive example is perhaps human culture, which is created through the aggregation and transmission of individual insights and opinions. However, while collective decision making seems to be universally beneficial in animals, it can have an adverse effect in humans, where the exchange of opinions can lead to the propagation of counter-factual rumors and can even give rise to the formation of radicalized groups. A deeper understanding of the process of collective opinion formation is needed if we are to determine the conditions under which it leads reliably to beneficial outcome. Significant progress is starting to be made on this problem, with several recent studies linking statistical physics models of opinion dynamics to experimental data \cite{Huepe2011,Couzin2011,Torok2013,deLuca2013}. For this effort to continue, the theoretical understanding of these systems must be expanded and systematic tools developed to reach analytical results. 

A paradigmatic model in this field is the adaptive voter model \cite{Ehrhardt2006,Holme2006}, describing a collection of individual agents whose opinions and social contacts may change over time. Agents hold one of two opinions, say $A$ and $B$, and are linked together by a sparse network of social interactions. The system evolves in time as follows: pairs of connected agents with opposing opinions are randomly chosen and either (i) the conflict is resolved by one agent adopting the opinion of the other, or (ii) one agent breaks the contact and forms a new link to a different agent. After a sufficiently long time the system reaches one of two types of absorbing state: a consensus state in which all agents hold the same opinion, or a fragmented state in which both opinions survive in disconnected groups \cite{Holme2006,Vazquez2008,Bohme2011,Durrett2012}. 

While the adaptive voter model has been explored in several recent studies, a larger body of previous work focuses on opinion dynamics on static networks (where the rewiring process does not occur and hence fragmentation is impossible). The main question addressed in these studies is the time taken for consensus to emerge \cite{Liggett1999,Cox1989,Sood2005}, which in general grows like $N^\mu$, where the exponent $\mu\leq1$ depends on the degree distribution of the underlying network \cite{Sood2005}. In contrast to this work, all the major analytical results in the adaptive networks literature have concerned the occurrence of fragmentation \cite{Holme2006,Vazquez2008,Bohme2011,Durrett2012}. The question of consensus time has so far been largely neglected, although some interesting results have been obtained via simulations \cite{Nardini2008} and heuristic scaling arguments \cite{Vazquez2008}.

\begin{figure}
\begin{center}
\includegraphics[width=0.36\textwidth, trim=50 0 50 0]{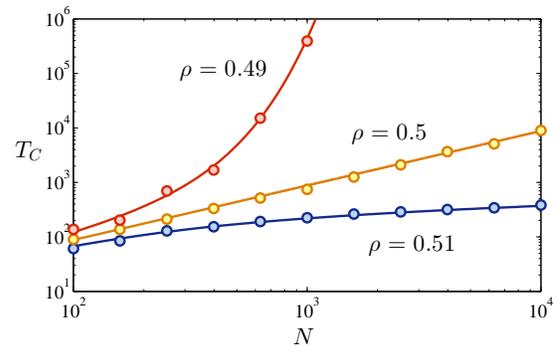} \\
\caption{(Color Online) Dynamical phase transition in the scaling behavior of the mean time to reach consensus $T_C$ as a function of network size $N$. At the critical value $\rho=1/2$ the growth of $T_C$ is linear, whilst being exponential for $\rho$ above the critical point, and logarithmic below. In each case circles give the average over 100 simulation runs with $k=10$ and $\phi=0.1$, while the solid lines show the result of the theory developed in the main text -- see equation (\ref{TC}).}
\label{exp_lin_log1}
\end{center}
\end{figure}

Here we describe a systematic and generally applicable analytical method to compute the time taken for consensus to emerge in the adaptive voter models. Furthermore, we show that when we allow pairs of agents to resolve conflicts by seeking another agent's opinion, this can either speed up or slow down the formation of consensus. This extension of the model exhibits a dynamical phase transition between exponential and logarithmic scaling laws, depending upon the probability of accepting the other agent's opinion. The original adaptive voter model is the critical case, exhibiting linear growth of consensus time.

Consider a network of $N$ nodes (agents) joined by a total of $K$ edges, which represent social interactions. Initially each agent is randomly assigned either opinion $A$ or $B$, and the edges are placed randomly. In each time step an edge $(i,j)$ is selected at random. If the focal edge connects agents that hold different opinions then it said to be active and the corresponding conflict is resolved in either of two ways. With probability $\phi$ the edge is rewired, with node $i$ cutting the edge and creating a new one to another node, selected at random from the set of all nodes holding the same opinion as $i$. If the edge is not rewired then a third opinion is sought: another node is selected at random from the rest of the graph and with probability $\rho$ both $i$ and $j$ adopt the opinion of the new node, otherwise, both $i$ and $j$ adopt the opposite opinion. This three-body interaction is a notable departure from the traditional voter model, although related systems have been studied previously \cite{Lambiotte2007}.

The parameter $\rho$ can be interpreted as a measure of social conformity \cite{Asch1956} and may range from $\approx1$ for a strongly conformist opinion formation process, to $\approx0$ for nonconformist processes. For $\rho=1/2$ the decision is not biased by the other agent, and the standard adaptive voter model is recovered. In most relevant contexts the stochastic response to another agent should not be interpreted as an actual consultation, but rather as an influence from the cultural `mean field' propagated by mainstream mass media.

As we will see, the model exhibits a dynamical phase transition as $\rho$ crosses the threshold $\rho_c=1/2$. Simulation results (Fig.~1) show that when agents accept the third opinion with probability $\rho>\rho_c$, the time to consensus only grows logarithmically with $N$, whereas in the case that $\rho<\rho_c$ it exponentially diverges. At precisely $\rho=\rho_c$ the original adaptive voter model is recovered, and we find a linear growth of the consensus time. 

For understanding the emergence of consensus in the model, we capture the dynamics of the system by a set of system-level variables that indicate the abundance of individual nodes and linked pairs of nodes with given opinion states. We denote the numbers of agents with each opinion by $[A]$ and $[B]$, and the numbers of edges between different agents by $[AA]$, $[AB]$ and $[BB]$. Because of the conservation laws for nodes, $[A]+[B]=N$, and edges \footnote{We note that the exact conservation of edges is assumed only for reasons of mathematical simplicity; similar results are found if this assumption is relaxed}, $[AA]+[AB]+[BB]=K$, the state of the system can be summarized by just three independent quantities: $X=[A]-[B]$, $Y=[AA]-[BB]$ and $Z=[AB]$. In analogy with spin glasses, the first two of these describe the magnetization of nodes and edges, while the third specifies the number of active edges and therefore controls the overall reaction rate of the system. 

Our analysis proceeds by deriving a closed set of rules for the stochastic dynamics of the variables $X$, $Y$ and $Z$, which approximate the evolution of the full network model. Introducing the system state vector $\bm{\Omega}=(X,Y,Z)^T$ ($T$ denotes transpose), we consider the effect of the four possible events which may occur in a given timestep: rewiring or updating of an $A$ or $B$ agent. For each, we write down the probability $r_i$ of occurrence in a given time step, and the average net change $\bm{s}_i$ to $\bm{\Omega}$ caused by the event. For example, an $AB$ link is chosen to be rewired to create and $AA$ link with probability $r_1=\phi Z /2K$ and the change to the system is $\bm{s}_1=(0,1,-1)^T$

Following \cite{Rogers2012}, we approximate the dynamics of $\bm{\Omega}$ by a Markov jump process known as the Pair-based Proxy (PBP). It is defined as follows: in each time step a jump vector $\bm{s}_i$ is chosen randomly with probability $r_i$, and the summary vector is updated by $\bm{\Omega}\mapsto\bm{\Omega}+\bm{s}_i$. The PBP represents a considerable reduction in complexity from the original adaptive network, and yet retains the essential stochastic nature of the system. 

In the limit of large network size, the PBP can be further reduced to a low-dimensional system of stochastic differential equations (SDEs). For simplicity, we package the update vectors into a stoichiometric matrix $\textbf{S}=(\bm{s}_1\cdots\bm{s}_4)$ and collect the event probabilities in a vector $\bm{r}=(r_1\cdots r_4)^T$. Defining the rescaled variables $x=X/N$, $y=Y/K$ and $z=Z/K$ and we apply Kurtz' theorem \cite{Kurtz1978} to obtain the following SDE:
\begin{equation}
\frac{\textrm{d}}{\textrm{d}t}\left(\begin{array}{c}x\\y\\z\end{array}\right)=\textbf{S}\bm{r}+\frac{1}{\sqrt{N}}\,\bm{\eta}(t)\,,
\label{SDE}
\end{equation}
where $\langle\eta_i(t)\eta_j(t')\rangle=\delta(t-t')B_{ij}\,,$ and the noise correlation matrix $\textbf{B}$ is given by
\begin{equation}
B_{ij}=\sum_{k}S_{ik}r_kS_{kj}\,.
\label{B}
\end{equation}
These equations can be written explicitly in terms of $x$, $y$, and $z$ if necessary. Let us consider the expression for the magnetization $x$ in detail, 
\begin{equation}
\frac{\textrm{d}x}{\textrm{d}t}=4(1-\phi)(\rho-1/2)xz+\frac{1}{\sqrt{N}}\eta_1(t)\,.
\end{equation}
The factor of $(\rho-1/2)x$ constitutes either positive or negative feedback depending on the value of $\rho$, which already suggests a transition in behavior around point $\rho_c=1/2$. The nonlinear interaction $xz$ shows that the dynamics require the presence of active edges, as well as an overall imbalance of opinions. 

\begin{figure}
\begin{center}
\includegraphics[width=0.4\textwidth, trim=60 0 80 0]{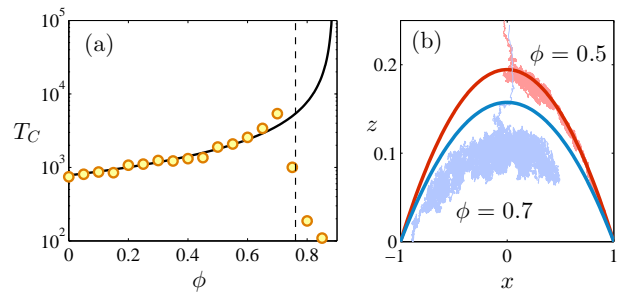} \\
\caption{(Color Online) (a) Dependence of consensus time $T_C$ on the rewiring rate $\phi$, when $\varepsilon=0$ and $k=10$. Orange circles show the average of 100 samples, the solid black line is the theoretical prediction (\ref{TC0}), while the dashed line indicates the point of the fragmentation transition, as derived in \cite{Bohme2011}. (b) Comparison between the slow manifold (\ref{fp}) and typical simulation trajectories for $\phi=0.5$ (red) and $\phi=0.7$ (blue). The discrepancy between simulations and theory in the case $\phi=0.7$ illustrates the breakdown of the pair-level approximation close to the fragmentation transition.}
\label{phi_and_parabolas}
\end{center}
\end{figure}

We make analytical progress by exploring the behavior of (\ref{SDE}) in the neighborhood of the transition, introducing $\varepsilon=\rho-1/2$. Let us first consider the case $\varepsilon=0$ in the deterministic limit $N\to\infty$. In this limit, the system (\ref{SDE}) possesses two manifolds of fixed points. The first is the plane $z=0$, which represents the state in which there are no active edges and thus the dynamics are frozen. These states are also absorbing states of the finite-size network model, corresponding to fragmentation. The second manifold of fixed points defines a parabola
\begin{equation}
y=x\,,\quad z=\frac{1}{2}(1-x^2)\frac{\phi_\ast-\phi}{1-\phi}\,,
\label{fp}
\end{equation}
where $\phi_\ast=(k-2)/(k-1)$ is the approximate critical rewiring rate for the fragmentation transition identified in \cite{Vazquez2008}. We note that Eq.~(\ref{fp}) is a pair-level approximation which becomes poor close to the fragmentation point; see \cite{Demirel2012} and Fig.~\ref{phi_and_parabolas}b. However, the question of consensus time concerns only values of $\phi$ below the transition, where we find the approximation to be sufficient for a large parameter range.

Local stability is governed by a linearization that is provided  by the  Jacobian matrix \textbf{J} of (\ref{SDE}). Computing the eigenvalues of the Jacobian on the parabola of active states we find 
$\lambda_1=0\,,$ $\lambda_2=2(\phi-\phi_\ast)/(2-\phi_\ast)\,,$ and $\lambda_3=\phi-\phi_\ast$. The corresponding eigenvectors are
\begin{equation}
\bm{v}_1=\left(\begin{array}{c}1\\1 \\x \mu_1\end{array}\right)\,,\bm{v}_2=\left(\begin{array}{c}0\\0 \\x\end{array}\right)\,,\bm{v}_3=\left(\begin{array}{c}0\\1 \\x\mu_2\end{array}\right)\,,
\end{equation}
where the constants are given by $\mu_1=-1+1/(k-1)(1-\phi)$ and $\mu_2=-1-2/(k-1)(k-2)(1-\phi)$. The second two eigenvalues are negative, and large in comparison to $\lambda_1=0$, meaning that trajectories close to the parabola collapse quickly in the directions of $\bm{v}_2$ and $\bm{v}_3$ (Fig.~\ref{phi_and_parabolas}b). This behavior, which will play a central role below, was previously noted in \cite{Vazquez2008} and is reminiscent of similar observations in the voter model on a static network \cite{Sood2005,Sood2008}.

Although the full stochastic system (\ref{SDE}) cannot be easily solved, we can derive an `effective' solvable one-dimensional system by restricting our attention to behavior in the neighborhood of the slow manifold, in analogy with \cite{Doering2012,Constable2013}. We reason as follows: in short time windows small Gaussian perturbations described by the noise correlation matrix defined in (\ref{B}) may move the system away from the slow manifold; for sufficiently small perturbations the net drift is then governed by the fast eigenvectors $\bm{v}_{2,3}$. We formalize this idea by fixing $y$ and $z$ to the values in (\ref{fp}) and replacing the noise matrix $\bm{B}$ with $\bm{PBP}^T$, where $\bm{P}$ is the linear projection whose range is spanned by $\bm{v}_1$ and kernel by $\bm{v}_{2,3}$. 

We thus obtain a reduced equation for motion on the manifold, in which $x$ is the only remaining variable,
\begin{equation}
\frac{\textrm{d}x}{\textrm{d}t}=2\varepsilon(\phi_\ast-\phi)x(1-x^2)+\sqrt{\frac{2(\phi_\ast-\phi)(1-x^2)}{N}}\,\xi(t)\,,
\label{1Deq}
\end{equation}
where $\xi$ is a standard Gaussian white noise variable. The picture we have now is as follows: from the initial condition the system state moves rapidly to the parabolic slow manifold (\ref{fp}), where it then drifts stochastically according to (\ref{1Deq}) until eventually reaching one of the absorbing consensus states at $x=\pm1$. 

We are interested in the mean waiting time before consensus is reached. Since our theory is one-dimensional, we follow \cite{Gardiner1985} to derive
\begin{equation}
T_C=\frac{N}{(\phi_\ast-\phi)}\int_0^1\int_0^y \frac{e^{\varepsilon N (x^2-y^2)}}{1-x^2}\,dx\,dy\,.
\label{TC}
\end{equation}
In the special case $\varepsilon=0$ the integral above can be computed easily to obtain
\begin{equation}
T_C=N\frac{\log(2)}{\phi_\ast-\phi}\,.
\label{TC0}
\end{equation}
In Fig.~\ref{phi_and_parabolas} we show a comparison between this prediction and the results of numerical simulations; the agreement is excellent for values of $\phi$ far from the fragmentation transition. This result validates the heuristic scaling argument presented in \cite{Vazquez2008}.

\begin{figure}
\begin{center}
\includegraphics[width=0.4\textwidth, trim=70 0 70 0]{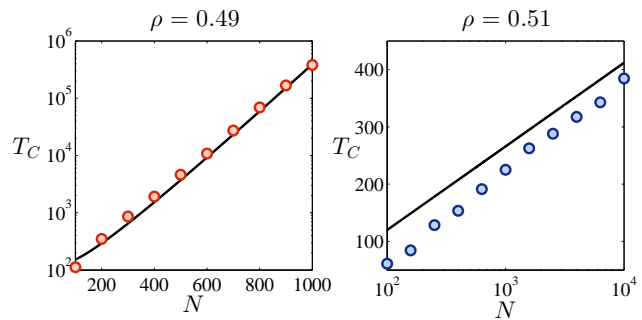} \\
\caption{(Color Online) Large $N$ scaling for $\rho$ either side of the critical value $\rho_c=1/2$. On both plots circles show the average consensus time for 100 simulation runs with $k=10$ and $\phi=0.1$. On the left the black line is given by Eq.~(\ref{TCE}), while on the right the slope is given by Eq.~(\ref{TCL}), whereas the intercept has been chosen manually for comparison.}
\label{exp_lin_log2}
\end{center}
\end{figure}

If $\rho<1/2$ then $\varepsilon$ is negative and the large $N$ asymptotic of (\ref{TC}) can be computed by Laplace's method as
\begin{equation}
T_C\approx\frac{e^{|\varepsilon|N}}{4(\phi_\ast-\phi)}\sqrt{\frac{\pi}{|\varepsilon|^3N}}\,.
\label{TCE}
\end{equation}
Alternatively, for $\rho>1/2$ the system (\ref{1Deq}) is deterministically unstable and thus the main contribution to $T_C$ comes from the initial symmetry breaking perturbation. In the neighborhood of the initial condition $x=0$, we have the linearized equation
\begin{equation}
\frac{\textrm{d}x}{\textrm{d}t}=2\varepsilon(\phi_\ast-\phi)x+\sqrt{\frac{2(\phi_\ast-\phi)}{N}}\,\xi(t)\,.
\end{equation}
Following \cite{Gardiner1985} again we find $\langle x^2\rangle=(e^{4\varepsilon(\phi_\ast-\phi)t}-1)/2\varepsilon N$, and thus the time taken for $x$ to reach a magnitude of order one is 
\begin{equation}
 T_C\sim\frac{\log{N}}{4\varepsilon(\phi_\ast-\phi)}\,.
 \label{TCL}
\end{equation}
A comparison of these predictions with numerical results (Fig.~\ref{exp_lin_log2}) shows an excellent agreement. These results establish a trichotomy between exponential, linear and logarithmic scaling laws, dependent on the parameter $\rho$. Note that the original adaptive voter model lies on the critical boundary between scaling regimes.

The above result suggests linear scaling of consensus time to be the exception rather than the rule, and likely to be destroyed by small changes in model specification. This is indeed the case, as can be seen by considering some other variants of the adaptive voter model. In some studies the target nodes in rewiring events are chosen randomly without regard to their opinion, \cite{Nardini2008,Durrett2012}. We refer to this as \textit{rewire-to-random}, as opposed to the \textit{rewire-to-same} scheme we considered above. 

The mechanism for choosing nodes to update may also be altered from the \textit{link update} rules we have used so far. Alternative model formulations use node update rules \cite{Nardini2008}, where one first chooses a node $i$ before selecting one of its neighbors $j$, and then in \emph{direct} node update $i$ copies $j$'s opinion, whereas in \emph{reverse} node update $j$ copies $i$. The corresponding models are the classical adaptive voter model (direct node update) and the adaptive invasion model (reverse node update). 

\begin{figure}
\begin{center}
\includegraphics[width=0.4\textwidth, trim=50 0 70 -3]{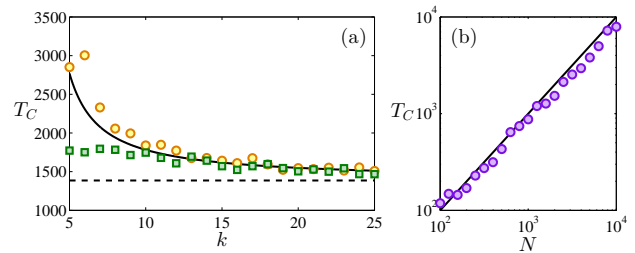} \\
\caption{(Color Online) Consensus time in other model specifications. (a) For link-update rules with moderate $k$, consensus time does not depend strongly on the rewiring rule. Theoretical predictions of Eq.~(\ref{TC0})(solid line) and the large $k$ limit (dashed line) are compared to simulation results averaged over 1000 networks for \textit{rewire-to-same} (orange circles) and \textit{rewire-to-random} (green squares). (b) Linear scaling for \textit{node-direct} updates in dense networks with $k=N/10$. The theoretical prediction of $T_C\sim N$ (solid line) is compared to simulation results averaged over 100 samples (purple circles). Parameters: $\phi=0.5$, $N=1000$.}
\label{esr_K}
\end{center}
\end{figure}

These changes in model specification result in different expressions for the event probability vector $\bm{r}$, however, the rest of the analysis may be repeated analogously. The results are summarized in Table~\ref{table} (see supplement for details). For link update rules and reasonably large $k$, we find that the choice of rewiring rule does not change the typical time to reach consensus. This effect is demonstrated numerically in Fig.~\ref{esr_K}a. However, using node update rules a range of scaling behaviors are possible. For example, the growth of $T_C$ is slightly slower than exponential in the case of node reverse and rewire-to-random. We can test this prediction by considering dense networks in which the average degree scales with the number of nodes according to $k=cN$. Here the theory predicts a return to linear growth, which is confirmed numerically in Fig.~\ref{esr_K}b. 

\begin{table}
\begin{center}
\begin{tabular}{p{20mm}|p{27mm}p{33mm}}
& Rewire-to-same & Rewire-to-random \\\hline
Edge & $T_C\sim N$ & $T_C\sim N$\\
Node-direct & $T_C\sim N$ & $T_C\sim k\log(Nk)$\\
Node-reverse & $T_C\sim N$ & $T_C\sim e^{N/k}\sqrt{k^3/N}$  
\end{tabular}
\end{center}
\caption{Summary of the scaling laws (in large $N$ and $k$) found for the mean time to consensus in various specifications of the adaptive voter model.}
\label{table}
\end{table}

In summary, we have formulated an analytical theory for the emergence of consensus in an extension of the adaptive voter model. By including the simple and sociologically plausible conflict resolution mechanism of seeking another opinion, we have shown how the formation of consensus may be enhanced or suppressed. This effect is manifested in trichotomy of scaling laws for consensus time, between exponential, linear and logarithmic. We also applied the proposed method to several other specifications of the model, showing how the previously observed sensitivity to model specification breaks down for highly connected networks. In all of the models investigated, consensus formation is driven by the intrinsic noise arising from the microscopic dynamical rules of the system. This noise is a universal feature of network models, being a result of the of the discrete nature of the individual interacting nodes and edges. We expect that the methods developed in this work will provide insights into emergent phenomena in other network systems.

\textit{Acknowledgements} TR thanks Alan McKane for brief but very useful discussions. 


%

\onecolumngrid
\large
\appendix
\newpage

\begin{center}
\begin{LARGE}\textbf{Supplementary material}\end{LARGE}\vspace{1cm}
\end{center}

\section{Derivation of the Pair Based Proxy}

In this section we describe the derivation of a low-dimensional Markov jump process which captures the essential features of the adaptive network model. The method involves (i) choosing a small number of appropriate summary statistics (for example, the numbers of certain types of edge), (ii) enumerating the possible events that bring about a change to these quantities, and (iii) computing the probability of these events and the average change resulting.

First, we recall the model definition. Consider a network of $N$ nodes and $K$ edges in which both the edges and the node opinions are initially assigned uniformly at random. In each time step an edge $(i,j)$ is selected at random. If the edge is active (that is, the end points have different opinions) then with probability $\phi$ the edge is rewired, with node $i$ cutting the edge and creating a new one to another node, selected at random, but with the same opinion as $i$. If the edge is not rewired then a second opinion is sought: another node is selected at random from the rest of the graph and with probability $\rho$ both $i$ and $j$ adopt the opinion of the new node, otherwise, $i$ and $j$ both take the opposite opinion. 

The state of the system at a given time is summarized by the state vector $\bm{\Omega}$, whose components are $X=[A]-[B]$, $Y=[AA]-[BB]$ and $Z=[AB]$. There are four basic events which can alter the state of the system, which we will consider in turn. First, a node in state $A$ may rewire one of its edges from a $B$ to an $A$. This results in $[AA]$ increasing by one, and $[AB]$ decreasing by one. The changes to the state vector $\bm{\Omega}$ are captured by the vector 
\begin{equation}
\bm{s}_1=\left(\begin{array}{c}0\\1\\-1 \end{array}\right)\,.
\end{equation}
The probability of this event is given by
\begin{equation}
r_1=\frac{[AB]}{K}\phi\frac{1}{2}=\frac{Z\phi}{2K}\,.
\end{equation}
From left to right, the contributing factors are (i) the chance of selecting an $A-B$ edge from the $K$ available, (2) the probability of rewiring and (3) the probability of picking the $A$ node to rewire. With the same probability, a $B$ node may also choose to rewire, resulting an increment to $[BB]$ and decrement to $[AB]$. Similarly, the jump vector for this event is 
\begin{equation}
\bm{s}_2=\left(\begin{array}{c}0\\-1\\-1 \end{array}\right)\,.
\end{equation}

The other possible events are $A$ or $B$ nodes altering their states after seeking a second opinion. Recall that an $A$ node will change to $B$ with probability $\rho$ if the second opinion was $B$ or $(1-\rho)$ if it was $A$. Taking all the factors together, the probability per time step for a node to change its opinion is $(1-\phi)(\rho[B]+(1-\rho)[A])[AB]/NK$ for $A$ nodes and $(1-\phi)(\rho[A]+(1-\rho)[B])[AB]/NK$ for $B$ nodes. 

To quantify the typical change to the system resulting from such an event requires some thought. Suppose an $A$ node copies the opinion of one of its $B$ neighbors; we can immediately deduce that $[B]$ and $[BB]$ will increase by one, while $[A]$ and $[AB]$ will decrease by one. In addition, we must consider the other neighbors of the node whose opinion changed. On average an $A$ node has $$k_A=\frac{2[AA]+[AB]}{[A]}$$ neighbors, of which we expect $2[AA]/[A]$ to be of type $A$ and $[AB]/[A]$ to be of type $B$. The node in question already has one neighbor that we know about, so the average number of additional $A$ neighbors it has is $(1-1/k_A)2[AA]/[A]$ and $B$ neighbors is $(1-1/k_A)[AB]/[A]$.

Putting this information together, we arrive at the following vector describing the typical change to $\bm{\Omega}$ brought about by an $A$ node changing its opinion:
\begin{equation}
\bm{s}_3=\left(\begin{array}{c}-2\\-k_A\\-1+\left(1-\frac{1}{k_A}\right)\frac{2[AA]-[AB]}{[A]}\end{array}\right)\,.
\end{equation}
To write $\bm{s}_3$ in terms of $X,Y$ and $Z$, we note the following:
\begin{equation}
\begin{split}
&[A]=\frac{1}{2}(N+X)\,,\quad[B]=\frac{1}{2}(N-X)\,,\\
&[AA]=\frac{1}{2}(K+Y-Z)\,,\quad[AB]=Z\,,\quad[BB]=\frac{1}{2}(K-Y-Z)\,,\\
&k_A=2\,\frac{K+Y}{N+X}\,,\quad k_B=2\frac{K-Y}{N-X}\,.
\end{split}
\end{equation}
Thus
\begin{equation}
\bm{s}_3=\left(\begin{array}{c}-2\\-2\,\frac{K+Y}{N+X}\\-1+(K+Y-2Z)\left(\frac{2}{N+X}-\frac{1}{K+Y}\right)\end{array}\right)\,.
\end{equation}

Finally, for $B$ nodes changing their opinion, the update vector $\bm{s}_4$ is obtained in exactly the same way:

\begin{equation}
\bm{s}_4=\left(\begin{array}{c}2\\k_B\\-1+\left(1-\frac{1}{k_B}\right)\frac{2[BB]-[AB]}{[B]} \end{array}\right)=\left(\begin{array}{c}2\\2\frac{K-Y}{N-X}\\-1+(K-Y-2Z)\left(\frac{2}{N-X}-\frac{1}{K-Y}\right) \end{array}\right)\,.
\end{equation}
There is an implicit assumption in the above derivation. When we estimated the number of additional $A$ and $B$ neighbors belonging to the updated $A$ node, we used the statistics of \emph{typical} $A$ nodes. However, we know that the node in question already has at least one $B$ neighbor; in certain circumstances this fact can mean that the node is actually rather unusual (for example if there are not may $AB$ edges). This assumption is essentially equivalent to the pair-approximation, which is regularly employed in studies of adaptive networks. The approximation is uncontrolled in the sense that there is no rigorous bound on the error committed, however, it is known to work well in the voter model provided $k$ is not too small and $\phi$ is not too large \cite{Demirel2012}.

For notational convenience, we package the update vectors into a stoichiometric matrix $S=(\bm{s}_1\,:\,\bm{s}_2\,:\,\bm{s}_3\,:\,\bm{s}_4)$, and collect the event probabilities in a vector 
\begin{equation}
\bm{r}=\frac{[AB]}{K}\left(\begin{array}{c}\phi/2\\\phi/2\\(1-\phi)(\rho[B]+(1-\rho)[A])/N\\(1-\phi)(\rho[A]+(1-\rho)[B])/N\end{array}\right)=\frac{z}{2}\left(\begin{array}{c}\phi\\\phi\\(1-\phi)(1+(1-2\rho)x)\\(1-\phi)(1-(1-2\rho)x)\end{array}\right)\,.
\end{equation}
Here we have introduced the rescaled variables $x=X/N$ $y=Y/K$ and $z=Z/K$. For $N\gg1$, application of Kurtz' theorem provides the mesoscopic equations
\begin{equation}
\frac{d}{dt}\left(\begin{array}{c}x\\y\\z\end{array}\right)=S\bm{r}+\frac{1}{\sqrt{N}}\,\bm{\eta}(t)\,,
\end{equation}
where
\begin{equation}
S=\left(\begin{array}{cccc}0&0&-2&2\\1&-1&-k\left(\frac{1+y}{1+x}\right)&k\left(\frac{1-y}{1-x}\right)\\-1&-1&k\left(\frac{1+y-2z}{1+x}\right)-2\left(\frac{1+y-z}{1+y}\right)&k\left(\frac{1-y-2z}{1-x}\right)-2\left(\frac{1-y-z}{1-y}\right)\end{array}\right)\,,
\end{equation}
and
\begin{equation}
\langle\eta_i(t)\eta_j(t')\rangle=\delta(t-t')B_{ij}\,,
\end{equation}
with the noise correlation matrix $B$ is given by
\begin{equation}
\begin{split}
B_{ij}&=\sum_{k}S_{ik}r_kS_{kj}\,.
\end{split}
\label{B2}
\end{equation}

\section{Slow manifold projection}
As explained in the main text, we are interested in the behavior of the model when $\rho\approx1/2$. We begin with an analysis of the case that $\rho$ is exactly equal to $1/2$. In this case the equations describing the system are, at last, short enough to be comprehensible:
\begin{equation}
\begin{split}
&\frac{dx}{dt}=\frac{1}{\sqrt{N}}\eta_x(t)\\
&\frac{dy}{dt}=2z(1-\phi)\frac{x-y}{1-x^2}+\frac{1}{\sqrt{N}}\eta_y(t)\\
&\frac{dz}{dt}=\frac{2z}{k}\left(\phi-2+\frac{2z(1-\phi)}{1-x^2}+\frac{k(1-\phi)(1-xy-2z)}{1-y^2}\right)+\frac{1}{\sqrt{N}}\eta_z(t)\,,
\end{split}
\end{equation}
with noise correlation matrix
\begin{equation}
\end{equation}
In the deterministic limit $N\to\infty$ the system possesses a pair of slow manifolds of fixed points: the plane $z=0$ and the parabola 
\begin{equation}
y=x\,,\quad z=\frac{1}{2}(1-x^2)\frac{\phi_\ast-\phi}{1-\phi}\,,
\label{fp2}
\end{equation}
where $\phi_\ast=(k-2)/(k-1)$. Whilst every point of the plane $z=0$ is unstable, the one-dimensional manifold described in (\ref{fp2}) is attractive. We compute the Jacobian matrix at a point $(x,y,z)$ chosen from the parabola:
\begin{equation}
J=\left(\begin{array}{ccc}0&0&0\\(\phi_\ast-\phi)&(\phi-\phi_\ast)&0\\\frac{x(1+\phi-2\phi_\ast)(\phi_\ast-\phi)}{1-\phi}&\frac{x(2(\phi_\ast-\phi-k(1-\phi))(\phi_\ast-\phi)}{k(1-\phi)}&\frac{2(\phi-\phi_\ast)}{2-\phi_\ast}\end{array}\right)
\end{equation}
since it is lower-triangular, the eigenvalues of $J$ may be read from the diagonal:
\begin{equation}
\lambda_1=0\,,\quad\lambda_2=\phi-\phi_\ast\,,\quad\lambda_3=\frac{2(\phi-\phi_\ast)}{2-\phi_\ast}\,.
\end{equation}
The corresponding eigenvectors are
\begin{equation}
\bm{v}_1=\left(\begin{array}{c}1\\1 \\x \mu_1\end{array}\right)\,,\bm{v}_2=\left(\begin{array}{c}0\\1 \\x\mu_2\end{array}\right)\,,\bm{v}_3=\left(\begin{array}{c}0\\0 \\1\end{array}\right)\,,
\end{equation}
where the constants are given by 
\begin{equation}
\begin{split}
&\mu_1=-1+\frac{1}{(k-1)(1-\phi)}\\
&\mu_2=-1-\frac{2}{(k-1)(k-2)(1-\phi)}\,.
\end{split}
\end{equation}
The second two eigenvalues are negative, and large in comparison to $\lambda_1=0$, meaning that trajectories close to the parabola collapse quickly in the directions of $\bm{v}_2$ and $\bm{v}_3$. 

\begin{figure}
\begin{center}
\includegraphics[width=0.5\textwidth, trim=0 50 0 -10]{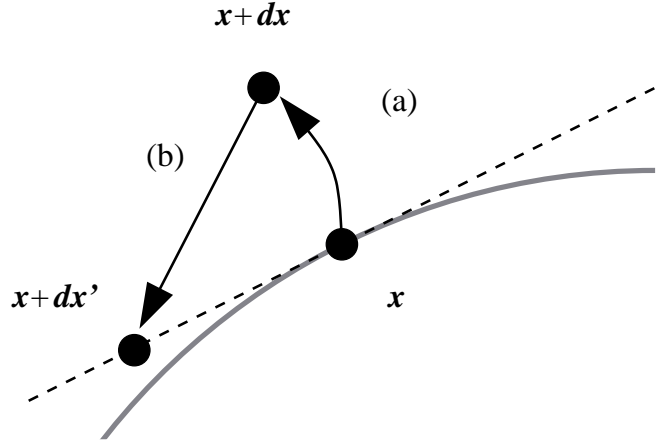} \\
\caption{Illustration of the projection.}
\label{projection}
\end{center}
\end{figure}

We intend to make this intuition formal by writing a reduced one-dimensional equation for motion along the slow manifold. We reason as follows: the full stochastic equations (\ref{SDE}) can be thought of as governing the random motion of a particle $\bm{x}$, whose coordinates tell us the state of the network model. Suppose that at time $t$ the particle is sitting on the slow manifold (\ref{fp2}). In a short time window, the noise terms in (\ref{SDE}) result in a random perturbation to $\bm{x}$ by the addition of a small `kick' $\bm{dx}$ (arrow (a) in Fig.~\ref{projection}); the entries of this vector are Gaussian random variables with covariance matrix $B$. Once the particle departs the slow manifold, the deterministic part of the dynamics becomes important again, causing a rapid decay along the direction of the fast eigenvectors (arrow (b) in Fig.~\ref{projection}).

We approximate the effective motion along the manifold by computing the projection of $\bm{x}+\bm{dx}$ in the direction of $\bm{v}_2$ and $\bm{v}_3$ onto the line through $\bm{x}$ tangent with $\bm{v}_1$. The final position can be written as $\bm{x}+\bm{dx'}$, where $\bm{x'}$ is a modified Gaussian random vector with covariance matrix $B'=PBP^T$. The projection matrix $P$ has the form $P=\bm{v}_1(\bm{u}^T\bm{v}_1)\bm{u}^T$, where $\bm{u}$ is any vector orthogonal to $\bm{v}_2$ and $\bm{v}_3$.

In the present case the $P$ has the simple form

\begin{equation}
\begin{split}
P=\left(\begin{array}{ccc}1&0&0\\1&0&0\\x\mu_1&0&0\end{array}\right)\,.
\end{split}
\end{equation}
The zeros in the second and third columns mean that, in the neighborhood of the slow manifold, it is the $x$ coordinate which dominates the dynamics. The reduced equation for $x$ alone is
\begin{equation}
\frac{dx}{dt}=\sqrt{\frac{2(\phi_\ast-\phi)(1-x^2)}{N}}\,\xi(t)\,,
\end{equation}
where $\xi$ is a standard Gaussian white noise variable. Applying the same method when $\rho\neq1/2$ yields the more general equation
\begin{equation}
\frac{dx}{dt}=(1-2\rho)(\phi_\ast-\phi)x(1-x^2)+\sqrt{\frac{2(\phi_\ast-\phi)(1-x^2)}{N}}\,\xi(t)\,.
\label{1Deq2}
\end{equation}

\section{Time to consensus}
In the space of $(x,y,z)$ variables, the consensus states are the points $(1,1,0)$ and $(-1,-1,0)$, corresponding to all nodes having opinion $A$ or $B$, respectively. These are also the endpoints of the parabolic slow manifold (\ref{fp2}). Moreover, the states $x=1$ and $x=-1$ are absorbing boundaries in the reduced equation (\ref{1Deq2}). The question of how long it takes the network to reach consensus is thus mapped onto the more accessible problem of determining the time until a particle obeying the SDE (\ref{1Deq2}) is absorbed. This problem is solved by textbook \cite{Gardiner1985} methods, which we now review. 

Consider a general one-dimensional It\={o} SDE
\begin{equation}
\frac{dx}{dt}=f(x)+g(x)\eta(t)\,.
\end{equation}
We write $T(x)$ for the expected time taken for a particle to hit either $-1$ or $1$, given that it is at position $x$ at time $t=0$. It can be shown that $T(x)$ satisfies the second-order ordinary differential equation
\begin{equation}
f(x)\frac{\partial}{\partial x}T(x)+\frac{1}{2}g(x)^2\frac{\partial^2}{\partial x^2}T(x)=-1\,,
\end{equation}
with boundary conditions
\begin{equation}
T(-1)=T(1)=0\,.
\end{equation}
This equation is solved with the introduction of the integrating factor $\psi(x)=e^{2\int_{-1}^x\frac{f(y)}{g(y)^2}dy}$, the general result is
\begin{equation}
T(x)=2\,\frac{\left(\int_{-1}^x\frac{1}{\psi(y)}\scriptstyle dy\right)\left(\int_{x}^1\frac{1}{\psi(y)}\int_{-1}^y\frac{\psi(z)}{g(z)^2}\scriptstyle{dzdy}\right)-\left(\int_{x}^1\frac{1}{\psi(y)}\scriptstyle{dy}\right)\left(\int_{-1}^x\frac{1}{\psi(y)}\int_{-1}^y\frac{\psi(z)}{g(z)^2}\scriptstyle{dzdy}\right)}{\int_{-1}^1\frac{1}{\psi(y)}\scriptstyle{dy}}\,.
\label{general_T}
\end{equation}
In the special case that $\psi$ is even and $x=0$, the denominator cancels with the leading factors in the numerator, to give \begin{equation}
\begin{split}
T(0)&=\int_{0}^1\frac{1}{\psi(y)}\int_{-1}^y\frac{\psi(z)}{g(z)^2}{dzdy}-\int_{-1}^0\frac{1}{\psi(y)}\int_{-1}^y\frac{\psi(z)}{g(z)^2}{dzdy}\\
&=2\int_{0}^1\frac{1}{\psi(y)}\int_{0}^y\frac{\psi(z)}{g(z)^2}{dzdy}\,,
\end{split}
\end{equation}
where the second line comes from using the symmetry of $\psi$ about the origin to infer that the contribution to the integral from below the $y$ axis is equal to the contribution from above. Applying this formula with $f$ and $g$ taken from equation (\ref{1Deq}) yields
\begin{equation}
T(0)=\frac{N}{(\phi_\ast-\phi)}\int_0^1\int_0^y \frac{e^{\varepsilon N (z^2-y^2)}}{1-z^2}\,dz\,dy\,.
\label{TC2}
\end{equation}
A solution to this integral may be written using hypergeometric functions, however it is not especially enlightening. We will instead focus on the asymptotic scaling behavior as $N$ grows large. As discussed in the main text, we find three different regimes, depending on the value of $\varepsilon$.

In the simplest case of $\varepsilon=0$ we obtain the exact solution
\begin{equation}
T_C=\frac{N\log(2)}{(\phi_\ast-\phi)}\,.
\label{TC02}
\end{equation}

For negative $\varepsilon$ we observe that if $N$ is large, the integral over $z$ is almost completely dominated by the contribution from the neighborhood of $z=0$. The standard approach in a situation such as this is known as Laplace's method, or the Gaussian approximation: the integrand is well-approximated by a Gaussian curve centered at zero. Thus,
\begin{equation}
\begin{split}
T(0)&\approx\frac{N}{(\phi_\ast-\phi)}\int_0^1\int_0^y e^{\varepsilon N (z^2-y^2)}\,dz\,dy\\
&\approx\frac{N}{(\phi_\ast-\phi)}\int_0^1e^{-\varepsilon y^2}\int_0^\infty e^{\varepsilon N z^2}\,dz\,dy\\
&=\frac{1}{2(\phi_\ast-\phi)}\sqrt{\frac{\pi N}{|\varepsilon|}}\int_0^1e^{-\varepsilon y^2}\\
&\approx\frac{e^{|\varepsilon|N}}{4(\phi_\ast-\phi)}\sqrt{\frac{\pi}{|\varepsilon|^3N}}\,,
\end{split}\label{TCe}
\end{equation}
where the final line comes expanding the exponent $-\varepsilon N y^2$ around its maximum at $y=1$. 

For positive $\varepsilon$, the reverse of the above logic applies: integral over $z$ is dominated by the boundary $z=y$. Unfortunately, the contribution from the denominator diverges as $y\to 1$, meaning that the double-integral is not accessible by Laplace's method. We change strategies at this point and instead focus our attention on the dynamical equation itself. 

If $\varepsilon<0$ then (\ref{1Deq2}) is deterministically unstable at $x=0$, meaning that after an initial perturbation the system approaches consensus rapidly. In this case, the role of intrinsic noise is simply to provide the initial perturbation. The short-time evolution of a trajectory starting at zero is well-described by a linear approximation. Expanding to first order around $x=0$ we have
\begin{equation}
\frac{dx}{dt}=2\varepsilon(\phi_\ast-\phi)x+\sqrt{\frac{2(\phi_\ast-\phi)}{N}}\,\eta(t)\,.
\end{equation}
This is an Ornsein-Uhlenbeck process, which admits an exact solution. At general times, $x$ has a Gaussian distribution with variance 
\begin{equation}
\langle x^2\rangle=\frac{1}{4\varepsilon N}\left(e^{2\varepsilon(\phi_\ast-\phi)t}-1\right)\,.
\end{equation}
Before the deterministic part of the dynamics can take over and drive the system to consensus, we must wait for $x$ to reach a magnitude of order one. Putting $\langle x^2\rangle=\delta$ and solving for $t$ gives
\begin{equation}
t=\frac{\log(1+\varepsilon \delta N)}{4\varepsilon (\phi_\ast-\phi)}=\frac{\log(N)}{4\varepsilon (\phi_\ast-\phi)}+\mathcal{O}(1)\,.
\end{equation} 
\section{Alternative model specifications}
In the model studied in the previous sections, we chose edges at random to update and in a rewiring event the new node was chosen to have the same opinion as the initial node. Both of these choices are somewhat arbitrary, and may be changed. In the literature, three possible update rules have been considered:
\begin{description}
\item[Edge] An edge is chosen at random. If the edge is not rewired, then one endpoint copies the other, chosen at random.
\item[Node-direct] One node is chosen at random, the other node is selected at random from the neighbors of the first. If the edge is not rewired, then the first node copies the opinion of the second.
\item[Node-reverse] One node is chosen at random, the other node is selected at random from the neighbors of the first. If the edge is not rewired, then the second node copies the opinion of the first.
\end{description}
There are also two possible choices for rewiring scheme:
\begin{description}
\item[Rewire-to-same] The new node is chosen to have the same opinion as the rewiring node.
\item[Rewire-to-random] The new node may have any opinion.
\end{description}
The procedure developed for the ES (edge dynamics, rewire to same) specification of the model also applies to the other variants, as we now describe. For clarity, we are considering only the case $\rho=1/2$, that is, the original adaptive voter model. 

In each of these schemes the basic events which change the state of the system remain the same, and thus the stoichiometric matrix $S$ is unchanged. The rates, however, are different in each case. For edge dynamics, we have 
\begin{equation}
\bm{r}_{ES}=\frac{[AB]}{2K}\left(\begin{array}{c}\phi\\\phi\\(1-\phi)\\(1-\phi)\end{array}\right)\,,\quad \bm{r}_{ER}=\frac{[AB]}{2K}\left(\begin{array}{c}\phi[A]/N\\\phi[B]/N\\(1-\phi)\\(1-\phi)\end{array}\right)\,.
\end{equation}
The extra factors of $[A]/N$ and $[B]/N$ in the rewire-to-random case come from the probability of choosing a node of the same opinion to rewire to. For the node-direct (D) and node-reverse (R) with rewire-to-same, the rate vectors are
\begin{equation}
\bm{r}_{DS}=\frac{[AB]}{N}\left(\begin{array}{c}\phi/k_{A}\\\phi/k_{B}\\(1-\phi)/k_{A}\\(1-\phi)/k_{B}\end{array}\right)\,,\quad \bm{r}_{RS}=\frac{[AB]}{N}\left(\begin{array}{c}\phi/k_{B}\\\phi/k_{A}\\(1-\phi)/k_{B}\\(1-\phi)/k_{A}\end{array}\right)\,.
\end{equation}
The overall denominator is $N$ in this case since we are choosing from the population of nodes, rather than edges. The factors of $1/k_{A}$ and $1/1k_{B}$ are the result of choosing from the neighbors of a node; these factors are reversed in the node-reverse dynamics. Finally, for node dynamics with rewire-to-random we find
\begin{equation}
\bm{r}_{DR}=\frac{[AB]}{N}\left(\begin{array}{c}\phi[A]/Nk_{A}\\\phi[B]/Nk_{B}\\(1-\phi)/k_{A}\\(1-\phi)/k_{B}\end{array}\right)\,,\quad \bm{r}_{RR}=\frac{[AB]}{N}\left(\begin{array}{c}\phi[A]/Nk_{B}\\\phi[B]/Nk_{A}\\(1-\phi)/k_{B}\\(1-\phi)/k_{A}\end{array}\right)\,.
\end{equation}

How will these variations to the rules of the model affect the time taken to reach consensus? We begin by examining the choice of edge dynamics. From the work of the previous sections, we know that 
\begin{equation}
T_{ES}=\frac{N\log(2)}{(\phi_\ast-\phi)}\,.
\end{equation}
For rewire-to-random, the SDEs for $x,y$ and $z$ are
\begin{equation}
\begin{split}
&\frac{dx}{dt}=\frac{1}{\sqrt{N}}\eta_x(t)\\
&\frac{dy}{dt}=\frac{2z(1-\phi)(x-y)}{1-x^2}+\frac{x}{k}+\frac{1}{\sqrt{N}}\eta_y(t)\\
&\frac{dz}{dt}=\frac{2z(1-\phi)(1-xy-2z)}{1-x^2}+\frac{4z(1-\phi)}{k(1-y^2)}+\frac{3\phi-4}{k}+\frac{1}{\sqrt{N}}\eta_z(t)\,,
\end{split}
\end{equation}
where the noise matrix $B$ is given by the usual formula. There is once again a manifold of fixed points, although now it does not have such a simple form:
\begin{equation}
\begin{split}
&y=x-\frac{1-x^2}{2k(1-1/\phi)}\,,\\
&z=\frac{(1-y^2)\Big((4-3\phi)(1-x^2)-2k(1-\phi)(1-xy)\Big)}{4\Big(1-x^2-k(1-y^2)\Big)(1-\phi)}\,.
\end{split}
\end{equation}
From here the calculation proceeds exactly as in the case discussed in the previous sections: we compute the Jacobian, and project onto the slow manifold along the fast direction. The result is a one-dimensional approximation for motion along the manifold, namely\normalsize
\begin{equation}
\frac{dx}{dt}=\sqrt{\frac{\left(1-x^2\right) (2 k (1-\phi )+(1-x) x \phi ) (x (1+x) \phi -2 k (1-\phi )) \left(4-2 k (1-\phi )-\left(3-x^2\right) \phi \right)}{k \left(4 k^2 (1-\phi )^2-x^2 \left(1-x^2\right) \phi ^2+4 k (1-\phi ) \left(\left(1+x^2\right) \phi-1 \right)\right)}}\,\xi(t)\,,\end{equation}\large
where $\xi(t)$ is a Gaussian white noise variable. Exact computation of the time to absorption at $x=-1$ or $x=1$ is difficult, however, to first order in large $k$ we have
\begin{equation}
\frac{dx}{dt}\approx\sqrt{\frac{(1-x^2)}{N}\left(2(1-\phi)+\frac{2-(1-x^2)\phi}{k}\right)}\,\xi(t)\,.
\end{equation}
Applying (\ref{general_T}), we find
\begin{equation}
\begin{split}
T_{ER}&=\frac{N\log(2)}{(1-\phi)}+\frac{N(\log(2)+\phi/4)}{k(1-\phi)^2}+\mathcal{O}(k^{-2})\\
&=T_{ES}+\mathcal{O}(k^{-1})\,.
\end{split}
\end{equation}
For large $k$, then, the choice of rewiring scheme makes very little difference to the time taken to reach consensus. We test this conclusion numerically in Figure 4a of the main text. 

We move on now to study node dynamics with rewire-to-same. For node-direct, we have
\begin{equation}
\begin{split}
&\frac{dx}{dt}=\frac{2z(1-\phi)(y-x)}{1-y^2}+\eta_x(t)\\
&\frac{dy}{dt}=\frac{2z\phi(x-y)}{k(1-y^2)}+\eta_y(t)\\
&\frac{dz}{dt}=2z\left(\frac{4z(1-xy)(1-\phi)}{k(1-y^2)^2}-\frac{1+(1+2(k+1)z)(1-\phi)-xy(2-\phi)}{k(1-y^2)}+(1-\phi)\right)+\eta_z(t)\,.
\end{split}
\end{equation}
and for node-reverse 
\begin{equation}
\begin{split}
&\frac{dx}{dt}=\frac{2z(1-\phi)(x-y)}{1-y^2}+\eta_x(t)\\
&\frac{dy}{dt}=\frac{2z(x-y)(2k(1-\phi)(1-xy)-(1-x^2)\phi)}{k(1-x^2)(1-y^2)}+\eta_y(t)\\
&\frac{dz}{dt}=2z(1-\phi)\Bigg(\frac{1+(2z-1)(k-1)-\frac{2-\phi}{1-\phi}(1-xy)}{k(1-y^2)}\\
&\qquad\qquad\qquad\qquad\qquad+\frac{(1-xy)^2-2(2z-1)(1-xy)}{(1-x^2)(1-y^2)}-\frac{1}{1-x^2}\Bigg)+\eta_z(t)\,.
\end{split}
\end{equation}
Although these expressions are somewhat different from each other, and from those of the rewire-to-same model with edge dynamics, they have exactly the same parabola of fixed points as discussed above, namely 
\begin{equation}
y=x\,,\quad z=\frac{1}{2}(1-x^2)\frac{\phi_\ast-\phi}{1-\phi}\,.
\end{equation}
In both cases, the eigenvalues on the slow manifold are 
\begin{equation}
\lambda_1=0,\quad\lambda_2=(\phi-\phi_\ast)\left(1-\frac{\phi}{k(1-\phi)}\right)\,,\quad \lambda_3=\frac{2(\phi-\phi_\ast)}{2-\phi_\ast}\,.
\end{equation}
The details of the noise projection are slightly different for each version of the dynamics. For node-direct we find the reduced equation
\begin{equation}
\frac{dx}{dt}=\sqrt{2(1-x^2)(\phi_\ast-\phi)\left(1+\frac{\phi(1-\phi)}{(k(1-\phi)+\phi)^2}\right)}\,\xi(t)\,,
\end{equation}
and for node-reverse
\begin{equation}
\frac{dx}{dt}=\sqrt{2(1-x^2)(\phi_\ast-\phi)\left(1+\frac{\phi(1-\phi)}{(k(1-\phi)-\phi)^2}\right)}\,\xi(t)\,.
\end{equation}
Both of these are only slight modifications of the result for edge dynamics with rewire-to-same. The time to consensus is found in the usual way, giving
\begin{equation}
\begin{split}
T_{DS}=\frac{N\log(2)}{(\phi_\ast-\phi)(1+\frac{\phi(1-\phi)}{(k(1-\phi)+\phi)^2})}\,,\quad T_{RS}=\frac{N\log(2)}{(\phi_\ast-\phi)(1+\frac{\phi(1-\phi)}{(k(1-\phi)-\phi)^2})}\,.
\end{split}
\end{equation}
The conclusion we can draw from this exercise is that, under the rewire-to-same rule, the choice of edge or node dynamics has very little difference on the mean time to reach consensus. As we will now see, this picture is dramatically different for the rewire-to-random rule. 

For node-direct dynamics with random rewiring, the SDEs are
\begin{equation}
\begin{split}
&\frac{dx}{dt}=\frac{2z(1-\phi)(y-x)}{1-y^2}+\eta_x(t)\\
&\frac{dy}{dt}=\frac{z\phi(x(2-xy)-y)}{k(1-y^2)}+\eta_y(t)\\
&\frac{dz}{dt}=\frac{z\phi(2xy-1-x^2)}{k(1-y^2)}+2z(1-\phi)\left(1+\frac{2(1-xy+(k+1)z)}{k(1-y^2)}+\frac{4(z-xyz)}{k(1-y^2)^2}\right)+\eta_z(t)\,,
\end{split}
\label{NDR}
\end{equation}
and for node-reverse
\begin{equation}
\begin{split}
&\frac{dx}{dt}=\frac{2z(1-\phi)(x-y)}{1-y^2}+\eta_x(t)\\
&\frac{dy}{dt}=\frac{\phi z y(1-x^2)}{k(1-y^2)}+\frac{4(1-\phi)(y-x)(1-xy)}{(1-x^2)(1-y^2)}+\eta_y(t)\\
&\frac{dz}{dt}=\frac{\phi z(x^2-1)}{k(1-y^2)}+2z(1-\phi)\Bigg(1-\frac{2}{1-x^2}+\frac{2((k+1)(z-1)+xy)}{k(1-y^2)}\\
&\hspace{95mm}+\frac{4(1-xy)(1-z)}{(1-x^2)(1-y^2)}\Bigg)+\eta_z(t)\,.
\end{split}
\label{NRR}0
\end{equation}
The behaviors of these versions of the model are fundamentally different from those previously considered: they do not posses a non-trivial manifold of fixed points. Instead, the fixed points for both (\ref{NDR}) and (\ref{NRR}) are the plane $z_\ast=0$, and the point 
\begin{equation}
x_\ast=0\,,\quad y_\ast=0\,,\quad z_\ast=\frac{\phi_\star-\phi}{2(1-\phi)(2-\phi_\star)}\,,
\end{equation}
where $\phi_\star=(2k-4)/(2k-3)$. The time to reach consensus in each of these systems is dominated by the behavior in the neighborhood of this fixed point.
\begin{equation}
\begin{split}
&\lambda_1=-\frac{2(\phi_\star-\phi)}{4-3\phi_\star}\\
&\lambda_{2,3}=-\frac{z_\ast}{2k}\left(2k(1-\phi)+\phi\pm\sqrt{\Big.\big(2k(1-\phi)+\phi\big)^2+8k(1-\phi)\phi}\right)\,.
\end{split}
\end{equation}
Notice that for all positive $k$ and $\phi\in[0,1]$ we have $$2k(1-\phi)+\phi<\sqrt{\Big.\big(2k(1-\phi)+\phi\big)^2+8k(1-\phi)\phi}\,.$$ So, for $\phi<\phi_\star$ we have that $\lambda_1$ and $\lambda_2$ are real and negative, but $\lambda_3$ is positive, meaning that the fixed point is unstable in the direction of the eigenvector $$\bm{v}_3=\left(\begin{array}{c}2k(1-\phi)-\phi-\sqrt{\Big.\big(2k(1-\phi)+\phi\big)^2+8k(1-\phi)\phi}\\4\phi\\0\end{array}\right)\,.$$
Typically, however, this instability is weak. Expanding in large $k$, we find
\begin{equation}
\begin{split}
&\lambda_1\approx-2(1-\phi)\,,\quad\lambda_2\approx-(1-\phi)\,,\quad\lambda_3\approx\frac{\phi}{2k}\,.
\end{split}
\end{equation}
For moderate values of $k$ then, typically trajectories collapse quickly towards $(x_\ast,y_\ast,z_\ast)$ in the direction of $\bm{v}_{1,2}$, whilst perturbations in the direction of $\bm{v}_3$ are weakly amplified. We exploit this observation to gain analytical traction by once again projecting trajectories, this time onto the manifold $\{(x_\ast,y_\ast,z_\ast)+s\bm{v}_3\,:\,s\in\mathbb{R}\}$. The resulting one-dimensional model is
\begin{equation}
\frac{dw}{dt}=\lambda_3 w + b\,\xi(t)\,,
\end{equation}
where $w$ describes the distance of the system from the fixed point, along the direction $\bm{v}_3$, $b$ is the amplitude of the projection of the noise matrix in the direction $\bm{v}_3$, and $\xi(t)$ is a standard Gaussian white noise. In the present case we have
\begin{equation}
b=\frac{1}{k}\sqrt{\frac{\phi(2-\phi)}{2(1-\phi)}}+\mathcal{O}(k^{-3/2})\,.
\end{equation}
Since $\lambda_3$ is positive the mean square deviation of $w$ will grow with time, specifically,
\begin{equation}
\langle w^2\rangle=\frac{b^2}{4\lambda_3}(e^{2\lambda_3t}-1)\,.
\end{equation}
The time taken for $|w|$ to reach a magnitude of order one (i.e. the time it takes the system to break its symmetry and achieve consensus) thus scales like
\begin{equation}
T\sim \frac{1}{2\lambda_3}\log\left(1+\frac{4\lambda_3}{b^2}\right)\approx\frac{k}{\phi}\log\left(1+4Nk\frac{(1-\phi)}{(2-\phi)}\right)\,.
\end{equation}
For node-reverse dynamics, the eigenvalues at the fixed point are
\begin{equation}
\begin{split}
&\lambda_1=-\frac{2(\phi_\star-\phi)}{4-3\phi_\star}\\
&\lambda_{2,3}=-\frac{z_\ast}{2k}\left(2k(1-\phi)-\phi\pm\sqrt{\Big.\big(2k(1-\phi)-\phi\big)^2-8k(1-\phi)\phi}\right)\,.
\end{split}
\end{equation}
This time, $$2k(1-\phi)-\phi>\sqrt{\Big.\big(2k(1-\phi)-\phi\big)^2-8k(1-\phi)\phi}\,.$$ So, for $\phi<\phi_\star$ all three eigenvalues have negative real part and the fixed point is therefore stable. To reach consensus requires a sufficiently large noise-drive excitation for the system to escape the pull of this attractive fixed point; once again a linear theory will be sufficient to describe this process. Expanding for large $k$ we find
\begin{equation}
\begin{split}
&\lambda_1\approx-2(1-\phi)\,,\quad\lambda_2\approx-(1-\phi)\,,\quad\lambda_3\approx-\frac{\phi}{2k}\,,
\end{split}
\end{equation}
so we are again in the situation that collapse in the direction of the first two eigenvectors is very rapid. Projecting onto $\bm{v}_3$ as usual, we obtain the same linear stochastic equation as above:
\begin{equation}
\frac{dw}{dt}=\lambda_3 w + b\,\xi(t)\,.
\end{equation}
This time $\lambda_3<0$, so we may apply the same calculation as in (\ref{TCe}) to deduce
\begin{equation}
T_C\approx \frac{1}{2}\exp\left(-\frac{\lambda_3}{b^2}\right)\sqrt{-\frac{\pi b^2}{\lambda_3^3}}\approx\frac{1}{2}\exp\left(\frac{N\phi}{4k(1-\phi)}\right)\sqrt{\frac{16\pi k^3(1-\phi)}{N\phi^3}}\,.
\end{equation}

\end{document}